\documentclass[journal]{IEEEtran}
\usepackage{amsfonts, amsmath, amssymb, epsfig, graphicx}
\usepackage{booktabs,tabularx,adjustbox}
\usepackage{floatrow}
\usepackage{multirow}
\usepackage{setspace}
\usepackage{cite}
\usepackage[stable]{footmisc}

\begin{document}
\title{EEG Classification by factoring in Sensor Configuration}


\author{Lubna~Shibly~Mokatren,~\IEEEmembership{Student Member,~IEEE,}
        Rashid~Ansari,~\IEEEmembership{Fellow,~IEEE,}
        Ahmet~Enis~Cetin,~\IEEEmembership{Fellow,~IEEE,}
        Alex~D.~Leow,~\IEEEmembership{Member,~IEEE,}
        Olusola~Ajilore,~
        Heide~Klumpp,~\IEEEmembership{Member,~IEEE,}
        and~Fatos~T.Yarman~Vural,~\IEEEmembership{Senior Member,~IEEE}
\thanks{L. Shibly Mokatren, R. Ansari and A. Cetin are with the Department
of Electrical and Computer Engineering, University of Illinois at Chicago, Chicago}
\thanks{ A. D. Leow,O. Ajilore and H. Klumpp are Department of Psychiatry, University of Illinois at Chicago, Chicago}
\thanks{F. T.Yarman Vural was with Department of Computer Engineering, Middle East Technical University, Ankara, Turkey}}

\maketitle

\begin{abstract}
Electroencephalography (EEG) serves as an effective diagnostic tool for mental disorders and neurological abnormalities. Enhanced analysis and classification of EEG signals can help improve detection performance. A new approach is examined here for enhancing EEG classification performance by leveraging knowledge of spatial layout of EEG sensors. Performance of two classification models - model 1 that ignores the sensor layout and model 2 that factors it in - is investigated and found to achieve consistently higher detection accuracy. The analysis is based on the information content of these signals represented in two different ways: concatenation of the channels of the frequency bands and an image-like 2D representation of the EEG channel locations. Performance of these models is examined on two tasks, social anxiety disorder (SAD) detection, and emotion recognition using a dataset for emotion analysis using physiological signals (DEAP). We hypothesized that model 2 will significantly outperform model 1 and this was validated in our results as model 2 yielded $5$--$8\%$ higher accuracy in all machine learning algorithms investigated. Convolutional Neural Networks (CNN) provided the best performance far exceeding that of Support Vector Machine (SVM) and k-Nearest Neighbors (kNNs) algorithms.
\end{abstract}

\begin{IEEEkeywords}
machine learning, EEG, emotion recognition, and SAD.
\end{IEEEkeywords}

\IEEEpeerreviewmaketitle


\section{Introduction}\footnote{A preliminary version of this manuscript is presented in IEEE EMBS Conference on Neural Engineering \cite{mokatren2019eeg}}
\label{sec:introduction}
\IEEEPARstart{T}{he} use of electroencephalography (EEG) is a popular mechanism for diagnosing mental states and brain disorders. EEG records electrical patterns resulting from the brain activity using electrodes wired onto the scalp. EEG is among a variety of brain imaging methods usually used for diagnosis in research and medical areas, such as positron emission tomography (PET), computed tomography (CT), and functional magnetic resonance imaging (fMRI). However, EEG has key distinguishing attributes of excellent temporal resolution and cost-effectiveness compared with other methods \cite{sanei2013eeg}.
The EEG waveform is divided into five main frequency bands \cite{article}: Delta, Theta, Alpha, Beta, and Gamma waves.
EEG is a very popular, noninvasive monitoring method which plays an important role as a diagnostic tool in brain–computer interface (BCI) applications. It is used to evaluate various mental disorders, such as Alzheimer’s disease, strokes, migraine, sleep disorders, and Parkinson’s disease \cite{siuly2016significance}. However, the analysis process of EEG is not always accurate as the data are complex and degraded by noise and artifacts. Therefore, providing a new model to help improve the accuracy of EEG analysis is very crucial.
In the past, many classification algorithms were devised for using EEG data \cite{lotte2007review}, such as, linear discriminant analysis, neural networks, support vector machines (SVM), nonlinear bayesian classifiers, k Nearest-Neighbour (kNN), hidden markov model, combination of classifiers, and other algorithms for EEG-based BCI which are mostly based on machine learning \cite{lotte2018review}.
Of the many past studies on EEG classification, only a few have paid attention to spatial locations and configuration of the EEG channel sensors for the purpose of creating models that may achieve improved performance in the analysis tasks. Factoring in the sensor topology was a key driving factor in our research. We investigate two classification models - model 1 that ignores the sensor layout and model 2 that factors it in with different interpolation methods. We hypothesize that model 2 would outperform model 1. This was validated in our results as the performance of model 2 surpassed that of model 1 by yielding $5$--$8\%$ higher accuracy in all machine learning algorithms investigated. Convolutional Neural Networks (CNN) provided the best performance far exceeding that of Support Vector Machine (SVM) and k-Nearest Neighbors (kNNs) algorithms. To demonstrate the significance of EEG analysis, two major classification tasks are examined: Social Anxiety Disorder (SAD) detection, and emotion recognition using DEAP dataset, which are detailed next. 

\subsection{SAD task}  

SAD, the world's third largest mental health problem, affects about $7\%$ of the population \cite{TheSocialAnxietyAssociation:2017:Online}. It is characterized by fear of negative evaluations and avoidance of social situations \cite{leichsenring2017social}. The process of diagnosing SAD was officially recognized in 1980 by Diagnosis and Statistical Manual for Mental Disorders (version DSM-III). Over the years, the criteria evolved and are now described in the fifth edition of the manual (DSM-5) \cite{hofmann2017cognitive}. The effectiveness and reliability of the process of DSM-5 diagnosis are critical in accurate assessment of the underlying disorder \cite{kraemer2012dsm}. The use of EEG in SAD diagnosis has seen only limited study. Identifying SAD patients by visual detection of differences in the EEG signals is impractical. Automated methods of detecting SAD such as those based on machine learning algorithms are therefore adopted paving the way for potentially highly accurate diagnosis, better connectivity analysis, and improved understanding of treatment responses in SAD \cite{moscovitch2011frontal}.

\subsection{Emotion recognition task} 
The study of EEG-based emotion recognition is very popular in many fields such as psychology, neuroscience, and computer science. Emotions are a very important factor in correct interpretation of actions and play a crucial role in all-day communication. There exist numerous recent research studies about EEG-based emotion recognition systems. Some of these studies are presented in Section \ref{related}.
EEG-based emotion recognition task can be subject-dependent or subject-independent \cite{liu2014real}. In this paper, both subject-dependent and subject-independent approaches are investigated.

\section{RELATED WORK} \label{related} 
There exist several recent research studies that involve EEG-based emotion recognition systems. In this section, a review of a number of emotion-based recognition studies is provided, as DEAP dataset is publicly available. This allows us to have a better perspective of the differences between the methods and of the variability in their performances.
Piho and Tjahjadi \cite{piho2018mutual} investigated reduced EEG data of emotions using a mutual information-based adaptive windowing and achieved average accuracy of $89.61\%$ and $89.84\%$ for valence and arousal, respectively. Chao et al. \cite{chao2018recognition} integrated deep belief networks with glia chains learning framework using multichannel EEG data and achieved average accuracy of $76.83\%$ and $75.92\%$ for valence and arousal states classification, respectively. In their study, Xu J. et al. \cite{xu2019eeg} proposed a baseline strategy of using power spectral density feature extraction methods and CNNs, and obtained  $81.14\%$ and $77.69\%$ for valence and arousal, respectively. Another recent study on DEAP dataset was conducted by Ganapathy and Swaminathan \cite{ganapathy2019emotion} using electrodermal activity signals and multiscale deep CNNs to achieve a classification accuracy of $81.25\%$ and $83.75\%$ for valence and arousal, respectively. The studies stated before followed procedures that did not exploit the knowledge of EEG sensor configuration. In contrast, Li et al. \cite{li2017human} took the spatial configuration into account, and proposed emotion recognition method using EEG multidimensional feature images and hybrid deep neural networks, where the highest accuracy achieved for valence and arousal is $82\%$. Another study that utilizes the spatial topology is proposed by Chao et al. \cite{chao2019emotion}. Here, the authors suggested a deep learning framework (CapsNet) based on a multiband feature matrix of the EEG signal. Finally, a 3D convolutional neural network for EEG-based emotion recognition is presented by Wang et al. \cite{wang2018emotionet}, where the electrode topology and time domain information are considered in constructing the input. A summary of these research studies is listed in Table~\ref{tab:survey}, for valence and arousal.
While desperate studies considered the spatial configuration in the construction process of their models, the work presented in this paper proposes a significantly different method, which is described in Section \ref{method}. The models differ from each other with respect to input construction and classification techniques. This includes, but is not limited to, input dimensions, input representation, and interpolation method (if exists) for the missing electrodes locations.

\begin{table*}[!t]
 \caption{Survey of Recent Studies in Emotion Recognition Using DEAP Dataset}
\label{tab:survey}
\centering
\begin{tabular}{ccccl}
 \toprule
Study   &Classification Method  &Spatial configuration used &Valence Accuracy &Arousal Accuracy\\
 \midrule
 Piho and Tjahjadi \cite{piho2018mutual} &   SVM/kNN/NB &No& $89.61\%$ &$89.84$\\ 
 Chao et al. \cite{chao2018recognition} &DBN-GCs &No & $76.83\%$ &$75.92\%$\\
 Xu J. et al. \cite{xu2019eeg} &CNNs  &No &$81.14\%$ &$77.69\%$  \\
 Ganapathy and Swaminathan \cite{ganapathy2019emotion}  &multiscale deep CNNs & No &$81.25\%$ & $83.75\%$\\
  Li et al.\cite{li2017human}   &hybrid CNN and LSTM & Yes & $\leq 82\%$ &$\leq 82\%$\\
  Hao et al. \cite{chao2019emotion}   &CapsNet& Yes & $66.73\%$ &$68.28\%$\\ 
  Wang et al. \cite{wang2018emotionet}   &EmotioNet& Yes & $72.1\%$ &$73.3\%$\\ 
  Proposed method &CNN & Yes & $91.85\%$ &$91.06\%$\\
\bottomrule
\end{tabular}
\end{table*}

\section{METHOD} \label{method}

\subsection{EEG recording}
\subsubsection{SAD dataset}
In this study, the EEG dataset we used was obtained from the Department of Psychiatry at the University of Illinois at Chicago (UIC). The multi-channel EEG data were collected from two subject groups. The first consists of 32 SAD patients, and the second of 32 healthy subjects. The the brain activity was analysed in resting state, without any presentation of stimuli. A total of 34 electrodes were positioned over the scalp to record the signals. The sampling frequency of the digitized signals is 1024 Hz.

\subsubsection{DEAP dataset}
\begin{figure}[]
\centering
\includegraphics[width=8cm]{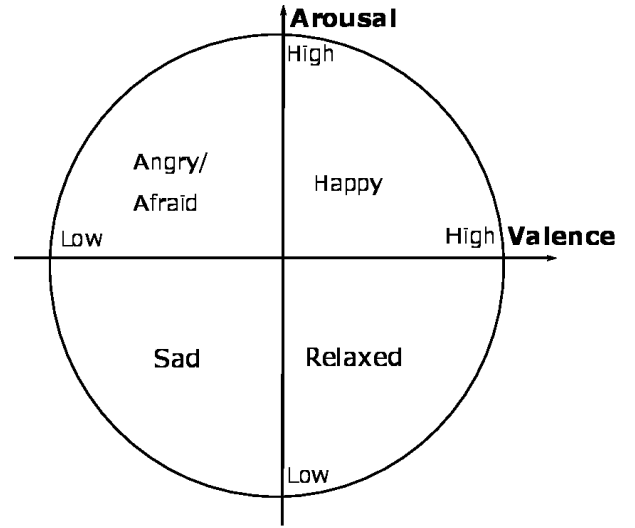}
\caption{Valence-Arousal space}
\label{fig1:Valence-Arousal space}
\end{figure}

The DEAP dataset is used for emotion analysis. Specifically, in this research, valence and arousal were assessed. DEAP is a publicly available EEG dataset \cite{koelstra2012deap} which contains signals from 32 participants. Each participant watched 40 one-minute long videos and evaluated themselves on the basis of four emotion states: arousal, valence, liking, and dominance on a scale of 1-9. The data are recorded over 32 channels on the scalp. For our analysis, a classification task of two output labels is considered. A rating value greater than or equal to 5 is set to 1 (aroused; pleasant), otherwise, it is set to 0 (relaxed; unpleasant). The 2D valence-arousal emotion space is shown in Fig. \ref{fig1:Valence-Arousal space}.

\subsection{Data Preprocessing}
EEG data are complex with high temporal resolution. The signals are easily contaminated with various artifacts and undesired noise, such as EMG artifacts, residual eye movements, and other muscle activities. Hence, preprocessing the data which includes artifacts removal, is very critical for proper analysis.  
\subsubsection{SAD data-set}
EEG data preprocessing stages are implemented using Fieldtrip and Brain Vision Analyzer (Brain Products, Gilching Germany) software. The data were converted to a linked mastoid reference and band-pass filtered.  Eye movements and ocular artifact corrections were performed \cite{gratton1983new},  and semi-automated artifact rejection of epochs where artifacts with voltage step higher than 50 $\mu V$ between samples were removed. Additional artifacts were removed by visual inspection.  The frequencies of interest are in the range of 0-48 Hz, covering the five different frequency bands ($\delta$,$\theta$,$\alpha$,$\beta$,$\gamma$).

\subsubsection{DEAP dataset}
The EEG data are downsampled to 128 Hz. EOG and EMG artifacts are removed. As delta waves usually corresponds to deeper sleep \cite{teplan2002fundamentals}, useful and informative data for emotion analysis are known to lie at 4-45 Hz frequency. Hence, a bandpass filter was applied leaving the first band, delta [0-4] Hz out of the analysis process. Eye artifacts were removed using a blind source separation technique, and the data were averaged to the common average reference (CAR) where the common average of the entire electrode is subtracted from a specific channel of interest, resulting in a zero-mean voltage distribution \cite{mcfarland1997spatial}. The data are segmented into 60-seconds trials for each video, and a 3 seconds pre-trial baseline is removed. 


\subsection{Data Analysis and Feature extraction} 
The constant switching of meta-stable states of neurons assembling during brain activity causes nonstationary phenomena to be manifested in EEG data, which consequently limits the reliability of the conventional analysis. In our data analysis, each channel signal is normalized and divided into N windows in which the data are assumed to be stationary. For each window, a wavelet packet decomposition (WPD) is applied to extract the five frequency bands, where the energy and entropy of these bands will be extracted as features. Based on the sampling frequency and statistical properties of the signals, multiple sizes of window segments were examined and a segment of size N=5 seconds and N=4 seconds provided better results for SAD and DEAP datasets, respectively. The subjects’ cognition and emotional states can be assessed by analyzing the EEG signal content in all frequency bands.

The energy and entropy of the content of each windowed segment is computed for the k frequency bands (k=5 for SAD, k=4 for DEAP) separately as described later in equations ``\eqref{eq:1}'' to ``\eqref{eq:4}''. The analysis is based on the energy and entropy content of these signals represented in two different ways: concatenation of the channels of the k frequency bands and an image-like 2D representation of the EEG channel locations. The latter method is discussed in Section \ref{IMGmethod}.


\subsection{Wavelet decomposition}

\begin{figure}[]
\centering
\includegraphics[width=8cm]{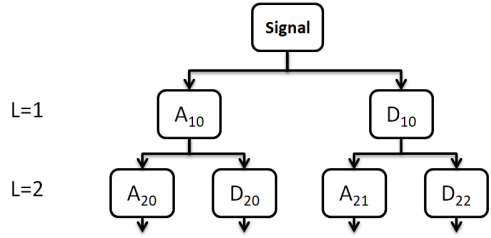}
\caption{Packet Wavelet Transform decomposition tree}
\label{pwdFig}
\end{figure}

After segmenting the data into multiple windows, wavelet packet decomposition (WPD) is applied to extract the EEG features. A packet wavelet transform decomposition tree with 2 levels is shown in Fig. \ref{pwdFig}. 
Since EEG signals are non-stationary, Fourier methods are not adequate enough for the time-frequency analysis of such signals. However, wavelet transforms can capture the local behavior of the signal, and can obtain both frequency and time information of transient non-stationary signals. Hence, they are more appropriate and preferable to use for EEG analysis and to decompose the signal into different bands \cite{adeli2003analysis}. According to \cite{kevric2017comparison}, features extracted from WPD subbands resulted in better classification accuracy than those extracted using DWT or empirical mode decomposition. For both datasets, WPD was used to extract the EEG frequency bands.  
In each level of WPD, the signal repeatedly passes through both low-pass and high-pass filters, followed by downsampling by 2. The output of each level is decomposed into detail (from the high-pass filter), and approximation (from the low-pass filter) coefficients.
For SAD dataset, the data are downsampled to 128 Hz and decomposed to multiple subbandes using wavelet packet decomposition with 4 levels and Daubechies 4 (\textit{db4}) mother wavelet. The approximate five frequency bands are chosen as follows: [0-4] Hz for $\delta$ band, [4-8] Hz for $\theta$, [8-12] Hz for $\alpha$, [12-32] Hz for $\beta$, and [32-48] Hz for $\gamma$. All five frequency bands are used for the analysis of SAD. This is illustrated in Table~\ref{tab:WPD}, which contains the decomposition coefficients of each subband.\\
In the case of the DEAP dataset, for each data segment, a 4-level WPD was applied to the input similarly to SAD dataset. Only 4 frequency bands, Gamma, Beta, Alpha and Theta are used for the emotion recognition task.

\begin{table}[!t]
 \caption{Wavelet Packet Decomposition for SAD}
\label{tab:WPD}
\begin{tabular}{ccl}
 \toprule
 Frequency band     &Frequency range(Hz)  &Decomposition level \\
 \midrule
 Gamma  &   $32-48$ Hz  & $A_{21}$\\
 Beta &$12-32$ Hz & $D_{41},D_{20}$ \\
 Alpha &$8-12$ Hz & $A_{41}$  \\
 Theta  &$4-8$ Hz & $D_{40}$  \\
 Delta  &$0-4$ Hz & $A_{40}$\\
\bottomrule
\end{tabular}
\end{table}

It should be noted that \textit{db4} wavelet is chosen due to its orthogonality and smoothing features, which are used for optimal detection of changes in EEG signal \cite{jahankhani2006eeg}. Moreover, according to \cite{murugappan2010classification}, extraction of EEG signal features using these wavelets could be more efficient as it has near-optimal time-frequency localization properties.

In both datasets, the energy and entropy content are both extracted as features. The mean wavelet energy $E_j$ of wavelet coefficients at decomposition level $j$ is defined as:
\begin{equation} \label{eq:1}
\begin{aligned}
\bar{E_j}=\frac{\sum_{k}|D_{j,k}|^{2}}{N_j}
\end{aligned}
\end{equation}
where $N_j$ is the number of wavelet coefficients at level $j$. The total energy is defined as
\begin{equation} \label{eq:2}
\begin{aligned}
E_{tot}=\sum_{j=1}^{N}\bar{E_j}
\end{aligned}
\end{equation}
and the relative wavelet energy is calculated as follows
\begin{equation} \label{eq:3}
\begin{aligned}
q_j=\frac{\bar{E_j}}{E_{tot}}
\end{aligned}
\end{equation}

The wavelet entropy is defined as 
\begin{equation}
\begin{aligned} \label{eq:4}
w_j=- q_jlogq_j
\end{aligned}
\end{equation}

\subsection{Image representation of the EEG data} \label{IMGmethod}
Data acquisition is performed by positioning \textit{M} electrodes over five areas on the scalp: Frontal (F), Central (C) Temporal (T), Parietal (P) and, Occipital (O), where \textit{M} is 34 for SAD data and 32 for DEAP data. It is believed that knowledge of the location of the channels can provide improved detection accuracy in the analysis of the data. To inspect this assumption, two main data models are examined using \textit{M} channels and \textit{B} extracted features. These features can be the energy of the frequency bands, or a combination of energy and entropy. $\textit{B} \in \{5;10\}$ for SAD, where \textit{B}=5 features when the energy of the 5 frequency bands is used for the analysis, and \textit{B}=10 when both energy and entropy are used. $\textit{B} \in \{4;8\}$ for DEAP as only 4 frequency bands were used in the analysis. 
In the first model, the \textit{M} channels of the \textit{B} features are concatenated by creating a \textit{M}$\times$\textit{B} feature matrix over each window, without accounting for the location of the channel electrodes (34$\times$\textit{B} for SAD, and 32$\times$\textit{B} for DEAP). For the second model, a 3D array of size \textit{K}$\times$\textit{K}$\times$\textit{B} is constructed consisting of a stack of B two-dimensional \textit{K}$\times$\textit{K} arrays each corresponding to a time snapshot of one of B features estimated from the electrode signals over a uniform \textit{K}$\times$\textit{K} grid. As explained later, we choose \textit{K}=15. In this model, the image pixels locations may not coincide with any of the M electrode locations and the B feature values over the \textit{K}$\times$\textit{K} grid are estimated by applying various interpolation techniques.

\begin{figure}[]
\centering
\includegraphics[width=8cm]{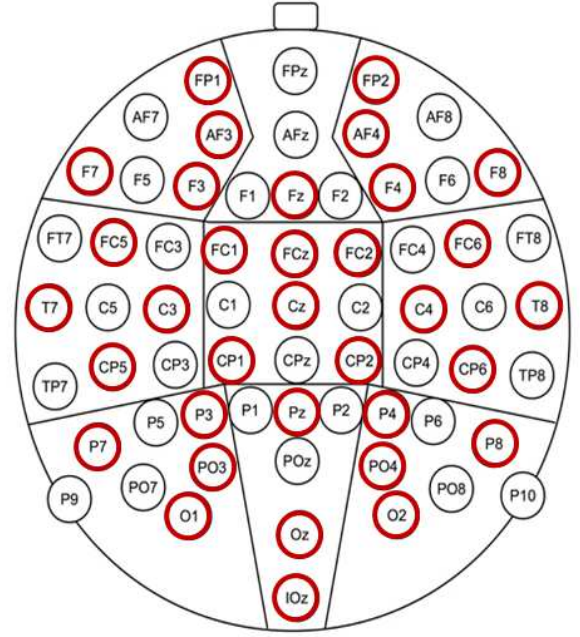}
\caption{Layout of 34 electrodes on scalp.}
\label{channelF}
\end{figure}

For both datasets, an image of size $15\times15$ was derived to construct an image-like representation of the channels layout. The \textit{M} channels are mapped to specific pixels in the image based on their locations. A layout of the channels' location for SAD data is shown in Fig. \ref{channelF}. The red circles represents the channels used in the recording of the data. Both datasets follow the international 10-20 System that allows EEG electrode placement to be standardized. To fill in the missing pixel feature values, the Inverse Distance Weighting (IDW) interpolation method was used \cite{eckstein1989evaluation}: for an interpolated value $e$ at point $x$, only the samples $u_i=u(x_i)$ with sensor locations $x_i$, $i=1,...n_i$, that are within a radius $d_{max}$ from the grid point $x$, are used in interpolating the value with the weighted average:

\begin{equation}
u(x)=\frac{\sum_{i=1}^{n_i} w_i(x)u_i}{\sum_{i=1}^{n_i}w_i(x)}
\end{equation}
If $d(x,x_i) \leq d_{max}$ is the distance between points $x$ and $x_i$, then $w_i(x)=\frac{1}{d(x,x_i)}$. At locations x where no sensors lie within a distance of $d_{max}$ i.e. "Border Points" (BP), the estimate is made with the nearest sensor value. The value of $d_{max}$ is empirically determined.

Other interpolation methods were also considered, such as IDW with zero values at border points (IDW with 0 BP), nearest neighbor interpolation, bilinear interpolation, and cubic b-spline interpolation \cite{lehmann1999survey}. However, they were all found to be inferior to the method mentioned above. This is further discussed in Section \ref{results}, where a summary of average performances of different interpolation methods in SAD, tested on the main CNN, is provided. 

It's worth noting that $K=15$ was selected empirically for constructing images of size \textit{K}$\times$\textit{K} as it gave higher accuracies when compared to other sizes in both datasets. Images of size \textit{K}$\times$\textit{K} were considered for the analysis where $K=\{10,15,20,25\}$. 
For SAD dataset, $K=15$ yielded the highest accuracy. For DEAP dataset, both $K=15$ and $K=20$ gave significantly good results. However, $K=15$ was chosen as the location is adequately captured without making the image size too large for computational load.

\section{EXPERIMENTS} \label{experiments}
After the data are collected and preprocessed, two models for each dataset are used as previously discussed.

SAD data: In the first model, the energy and entropy for the five frequency bands are calculated separately in each window of 5 seconds. Hence, for each window a feature matrix of dimensions 34$\times$\textit{B} is constructed. There are 34 channels or electrodes and 2 values of energy and entropy are extracted from each of the 5 frequency bands, yielding $\textit{B} \in \{5;10\}$. Each row in the matrix represents one channel.
The second model adopts the image representation technique of the EEG data, described earlier in Section \ref{IMGmethod}. A 3D array of 15$\times$15$\times$\textit{B} is constructed for each segment, where the third dimension comprises of the energy and entropy values of all the frequency bands.

DEAP data: The models are built in a manner similar to SAD models. However, the dimensions are different. For the first model, the energy and entropy for 4 frequency bands only are calculated separately in each window of 4 seconds. Hence, for each window an energy matrix of dimensions 32$\times$\textit{B} is constructed, as 32 channels were considered, $\textit{B} \in \{4;8\}$. In the second model, a 3D energy array of 15$\times$15$\times$\textit{B} is built for each window.
For both models in each dataset, each matrix is considered as a single sample for the training or testing data-set.

\subsection{Acquisition of training and testing EEG data}

\subsubsection{SAD dataset} To train the network, a stratified 8-fold cross-validation procedure is applied where the classifier is trained 8 separate times using a different fold for testing in each run. A total of 7 folds (56 subjects) are used for training and validation and the remaining fold (8 subjects) is used in the testing stage. 
In both models, for every window of size $N$, a feature matrix is constructed and considered as a single sample, where $N=5$ seconds. For each subject, multiple samples are gathered by sliding a moving window of size $N$ with no overlap of windows over all channels. The samples collected for the training, validation and testing sets do not overlap, i.e. different samples collected from specific subject cannot be used for training and testing at a certain trial.
Samples are labeled 1 if they belong to SAD patients, and 0 otherwise.
For every trial, each testing subject is evaluated as follows: 
\begin{equation}
prediction=\left\{ \begin{array}{cl} patient & \mbox{if}  ~x_i=\frac{p_i}{t_i} \geq Th  \\ control & \mbox{else}  \end{array}\right.
\end{equation}
where  $Th=0.5$, $p_i$ is number of samples classified as 1 for subject $i$, $t_i$ is total number of samples for subject $i$.

\subsubsection{DEAP dataset}A very important aspect in emotion recognition task is whether it is subject-dependent or independent. A subject-dependent task means partitioning the training and testing data from same participants. However, different samples taken from the same video cannot be used in both the training and testing data. Subject-independent task means that testing is performed on a group of participants which is different from the training group. According to \cite{soto2009emotion} there exists a physiological linkage with emotion recognition, which makes the recognition depend on age, culture, and gender. Both the structure of the training and testing data, and the physiological linkage make the performance accuracy in subject-independent task lower than subject-dependent. A summary of the comparison is provided in Section \ref{results}. 

A stratified 8-fold cross-validation is applied on the DEAP dataset. Again, 7 folds are used for training and validation, and the remaining fold is used for testing. The average accuracy is found after the classifier is trained 8 times. In the subject-independent task, the 7 folds correspond to 28 participants and the last fold corresponds to the 4 remaining participants. The window size taken to create a sample is $N=4$ seconds. For every 60 seconds of video, the samples are gathered from all 32 channels by sliding a window of size $N$ with $N/2$ overlap. 

The choice of the shift size for each dataset was made empirically as it yielded better results when compared with other window slides. The window shifts that are tested are $N/4$, $N/2$, $N$, and $3N/2$. Early stopping was applied by monitoring the validation loss to avoid overfitting.

The number of folds for cross-validation was chosen to be $k=8$ for both datasets after exploring various folds of sizes $k=\{5,8,10\}$. Fold of size $k=8$ was considered the best in terms of accuracy, computational cost, and being a divisor of the subjects size.

\subsection{Data Augmentation}
One of the key challenges in machine learning algorithms in general and deep neural networks specifically, is not having sufficient training data to properly perform a classification task \cite{lemley2017smart}. Training with small datasets might cause the model to be highly biased to the data in the training set, making the model perform poorly on the validation or testing set, as it cannot generalize what it learned to unseen samples. These models suffer from overfitting. Regularization, dropout, batch normalization, and data augmentations are some of the methods used to tackle the problem of overfitting \cite{wang2017effectiveness}.
Image augmentation technique is introduced to help improve the classification performance by creating more robust models with the ability to generalize. Data augmentation refers to artificially generating data by creating new samples to expand the training dataset. It is done by performing transformations on the original images while preserving the label which is invariant to certain variations.
In this work, horizontal and vertical shift augmentation is used to expand the training set. This translation is done by moving the image along the X or Y directions, specifically a shift of 1 or both 1 and 2 pixels is done on the image while preserving the image size. For model 1, images are shifted by a certain number of pixels to the right, left, up and down. The deleted rows or columns are simply replaced by the previous row or column respectively. For model 2, the shift is done over the first 2 dimensions only. The third dimension which corresponds to the energy or entropy of the frequency bands is filled according to the feature value it has at that specific location.

\subsection{CNN Network Structure} \label{cnnStruct}

In recent years, the use of deep learning solutions has become very popular in many applications. Deep learning-based methods have repeatedly shown improved performance compared with other classical machine learning algorithms on a wide variety of problems \cite{deng2014deep}. Specifically, CNNs have rapidly become a methodology of choice for image-related tasks, including, medical images processing \cite{krizhevsky2012imagenet}. CNNs are a specific kind of feedforward deep neural networks. Their architecture is characterized by arranging convolutional layers, pooling layers, and fully-connected layers. The input to the network is arrays of data containing energy and entropy values of subband signals which can be viewed as images. \\

\begin{figure*}[ht]
\centering
\includegraphics[width=15cm]{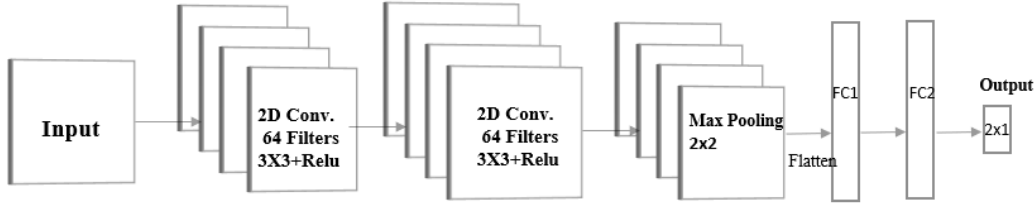}
\caption{Convolutional neural network structure.}
\label{cnnS}
\end{figure*}

A sequential model is built for SAD classification task as shown in Fig. \ref{cnnS}. The first layer in this model is a 2D convolution layer with kernel size 3$\times$3, 64 output filters, and ReLu activation over the outputs. It is then followed by a similar 2D convolution layer, and a max-pooling layer with pool size of 2. A dropout with rate=0.25 is performed. The input is then flattened and fed to a fully connected layer with 128 output dimension and ReLu activation. Another dropout is performed with rate=0.2 followed by a final fully connected layer with Softmax activation and output dimension that equals to 2 corresponding to the two labels in the recognition task. Dropout is a regularization method that is used to reduce over-fitting. A similar configuration is used for emotion recognition task, where the first layer is a 2D convolution layer with kernel size 3$\times$3, 32 output filters, followed by a max-pooling layer of size 2 $\times$ 2. Another convolution layers follows with 64 output filters and a max-pooling layer with pool size of 2. Dropout with rate=0.45 is performed. The output is then flattened and fed to a fully connected layer with Relu activation and 64 output dimension. Another dropout is performed with rate=0.25 followed by a fully connected layer with Softmax activation and output dimension that equals to 2.  
It should be noted that batch normalization was applied to the input from both datasets and the convolutional layers to reduce the internal covariate shift \cite{ioffe2015batch}. 
We started from a filter size of 3$\times$3 for the convolutional layers as the input has relatively small-sized dimension. The filter size, number of layers, and feature maps are further increased, but the structures specified above achieved better results.

\section{RESULTS} \label{results}

In the study, inputs are constructed based on two models. These models are investigated for their classification performance using different classifiers, different types of feature values, and two different datasets. Model 1 is constructed with no consideration of the spatial locations of EEG electrodes by simply concatenating the channels of the frequency bands. Model 2, on the other hand, factors in the electrode spatial configuration over the scalp.

The inputs to the classifiers of SAD and DEAP datasets are built using two groups of features: energy of the frequency bands, or a combination of energy and entropy.

\subsection{SAD dataset}
The main results for SAD dataset classification are presented using confusion matrices. In this case, accuracy is defined as the ability to correctly classify a subject. The analysis is subject-independent, and the confusion matrices for model 1 and model 2 can be seen in Table~\ref{tab:SAD1cm} and Table~\ref{tab:SAD2cm}, respectively. The top part of each table represents inputs built using energy of the frequency bands as features, and the bottom part represents inputs built using both energy and entropy. In the third column of each table, the first entry includes the number of actual SAD patients predicted as positive (patients), the number in the second entry is actual patients predicted as negative (healthy subjects), etc.
\begin{table}[!t]
 \caption{Confusion Matrix for SAD - Model 1}
\label{tab:SAD1cm}
\scalebox{0.5}{
\begin{adjustbox}{width=17cm}
\begin{tabular}{ccccll}
\toprule
\textbf{Model 1}&  &Actual Positive    & Actual Negative& \textbf{Accuracy} & \textbf{F1 Score} \\
\midrule
\midrule
\textbf{Energy} &Predicted Positive  & 26 & 6  &$81.25\%$ &$81.25\%$ \\
&Predicted Negative & 6     & 26  \\ \midrule
\textbf{Energy \& Entropy} &Predicted Positive & 28 & 6 &$84.38\%$ &$84.85\%$ \\
&Predicted Negative & 4     & 26  \\\bottomrule
\end{tabular}
\end{adjustbox}
}
\end{table}
\begin{table}[!t]
 \caption{Confusion Matrix for SAD - Model 2}
\label{tab:SAD2cm}
\scalebox{0.5}{
\begin{adjustbox}{width=17cm}
\begin{tabular}{ccccll}
\toprule
\textbf{Model 2}& & Actual Positive     &Actual Negative& \textbf{Accuracy} & \textbf{F1 Score} \\
\midrule
\midrule
\textbf{Energy} &Predicted Positive  & 29 & 4 &$89.06\%$ &$89.23\%$ \\
&Predicted Negative & 3     & 28  \\ \midrule
\textbf{Energy \& Entropy} &Predicted Positive  & 30 & 3 &$92.19\%$ &$92.31	\%$ \\
&Predicted Negative  & 2     & 29  \\\bottomrule
\end{tabular}
\end{adjustbox}}
\end{table}
We observe that regardless of the number of features used, the accuracy and F1-score for the proposed approach based on model 2 are higher. We conclude that model 2 that factors in the location of the electrodes provides a superior performance, achieving a maximum classification accuracy of $92.19\%$ compared with $84.38\%$ for model 1. Another important observation is that the number of features influences the accuracy, and it is higher when using both entropy and energy features. 


In this task, SVM with radial basis function (RBF) and kernel parameter $\sigma=0.4$ was considered as a classifier. The optimum values of the hyper-parameters were selected with a grid search method. In addition, k-NN classifiers with k=3 and k=5 were investigated, but were clearly outperformed by the proposed CNN network in terms of overall classification accuracy, as seen in Table~\ref{tab:SADavg} for data using energy and entropy features.

It should be noted that model 2 gave significantly higher accuracy than model 1, regardless of the classifier type or feature values. Establishing the importance of 2D representation of the spatial configuration of EEG sensors.


\begin{table}[!t]
 \caption{Average Performance on SAD Dataset(\%)}
\label{tab:SADavg}
\scalebox{0.5}{
\begin{adjustbox}{width=17cm}
\begin{tabular}{lllllc}
\toprule
  & Energy     & & &Energy \& Entropy \\
 \midrule\midrule
 Classifier  &Model 1 &Model 2 & &Model 1 &Model 2 \\ 
 \cmidrule{2-3}  \cmidrule{5-6} 
 kNN (k=5) & 70.31   & 73.43 &   &70.31 & 75      \\ 
kNN (k=3) & 70.31   & 76.56 && 71.87&   78.13     \\ 
SVM & 73.43   &  80.5    && 76.56&    81.25     \\ 
 CNN  & 81.25   &  89.06  & &84.38 & 92.19      \\ \bottomrule
\end{tabular}
\end{adjustbox}}
\end{table}

\subsection{DEAP dataset}
In this subsection, the binary valence and arousal states, low/high valence (LVHV) and low/high arousal (LAHA), are estimated. The classification is evaluated under two study cases, subject-dependent and subject-independent. The classifiers used for emotion recognition task are the CNN network described in Section \ref{cnnStruct}, SVM, kNN classifier with k=3 neighbors, and kNN with k=5. The average performance of model 1 and model 2 is evaluated using energy of decomposed frequency bands as features, or both energy and entropy.
Table~\ref{tab:DEAPsI} represents the performance of subject-independent task for valence and arousal recognition. Table~\ref{tab:DEAPsD} summarize the accuracies for subject-dependent task on valence and arousal. It is detected from the mentioned tables that model 2 outperforms model 1 in all cases, regardless of feature selection, classifier, or nature of the study case (subject dependent/independent). For instance, in subject-independent valence recognition, the highest average accuracy is $91.85\%$ for model 2 and only $84.77\%$ for model 1.

By comparing the average accuracies presented in the tables, it is included that including the entropy in the feature selection scheme has improved the performance of the classifiers. 

Another important observation is the noticeable differences in performance between the subject-dependent and subject-independent cases. The manner of construction of the training and testing data in these two cases makes the subject-dependent classifier achieve higher accuracies. However, the lack of generalization of such classifiers is the price one has to pay.

\begin{table}[!t]
 \caption{Subject-Independent Average Performance - DEAP(\%)}
\label{tab:DEAPsI}
\scalebox{0.5}{
\begin{adjustbox}{width=17cm}
\begin{tabular}{llllllll}
\toprule
   &   && Energy & &&Energy \&Entropy \\
   \cmidrule{4-5}  \cmidrule{7-8} 
  Emotion  & Classifier && Model 1& Model 2& & Model 1& Model 2\\ 
  \cmidrule{1-2}\cmidrule{4-5}  \cmidrule{7-8} 
 Valence   &kNN (k=3)  && 73.87 & 78.19&& 75.55& 81.65 \\
           & kNN (k=5)  &&  71.92 & 76.4&& 73.83& 80.73\\
           & SVM &&  75.36 & 81.52&& 78.13& 83.38\\ 
           & CNN  &&  82.36 & 87.64&& 84.77& \textbf{91.85}  \\ \midrule
 Arousal    & kNN (k=3)  && 71.93 & 76.06&& 73.8& 80.64 \\  
           & kNN (k=5) && 69.87 & 75.41&& 71.33& 77.58\\ 
            & SVM  && 74.16 & 77.06&& 75.82& 81.43\\ 
           & CNN  && 82.61 & 85.94&& 83.94& \textbf{91.06}\\ \bottomrule
\end{tabular}
\end{adjustbox}}
\end{table}

\begin{table}[htb]
 \caption{Subject-Dependent Average Performance - DEAP(\%)}
\label{tab:DEAPsD}
\scalebox{0.5}{
\begin{adjustbox}{width=17cm}
\begin{tabular}{llllllll}
\toprule
   &   && Energy & &&Energy \&Entropy \\
   \cmidrule{4-5}  \cmidrule{7-8} 
  Emotion  & Classifier && Model 1& Model 2& & Model 1& Model 2\\ 
  \cmidrule{1-2}\cmidrule{4-5}  \cmidrule{7-8} 
 Valence   &kNN (k=3) && 79.03 & 82.85&& 81.56& 85.93\\ 
           & kNN (k=5) && 76.45& 81.03&& 77.24& 81.92\\ 
           & SVM && 80.27 & 84.96&& 81.68& 85.31\\ 
           & CNN  && 84.63 & 88.97&& 88.02& \textbf{94.48}\\ \midrule
 Arousal  & kNN (k=3)&& 77.86 & 82.55&& 79.91& 84.76\\  
           & kNN (k=5) && 77.12 & 80.05&& 77.94& 82.05 \\ 
            & SVM && 76.34& 80.93&& 79.75& 83.04\\ 
           & CNN && 83.72 & 90.26&& 86.07& \textbf{93.66}\\ \bottomrule
\end{tabular}
\end{adjustbox}}
\end{table}


Due to the non-normal distribution of the classification accuracies, Wilcoxon Signed Rank test was conducted to investigate the statistical significance. We define the null hypothesis to be that the two models generally behave the same and the alternative hypothesis to be that model 2 has improved performance. A significance level of 0.05 was considered.
Table~\ref{tab:stat} summarizes the statistical significance of classification accuracy differences between model 1 and model 2 for subject-dependent approach. The analysis is performed for valence and arousal classification using energy and entropy features for all four classifiers used to evaluate the performance. The results indicate that model 2 yielded significantly higher accuracy than that of model 1 in almost all cases.

\begin{table}[!t]\centering
\caption{Statistical analysis of classification accuracy differences between the compared models for different classifiers}
\label{tab:stat}
\begin{tabular}{@{}llllc@{}}\toprule
& \multicolumn{4}{c}{Classifier} \\
\cmidrule{2-5} 
& kNN(k=3) & kNN(k=5)& SVM & CNN\\ \midrule
Arousal & p=0.002 & p=0.0078 & p=0.000& p=0.012\\\midrule
Valence & p=0.063& p=0.026& p=0.000& p=0.001\\
\bottomrule
\end{tabular}
\end{table}


The EEG channels' locations are represented as an image of size \textit{K}$\times$\textit{K} in model 2. There are 34 channels in SAD dataset, and 32 in DEAP. To fill out the empty pixels, a few interpolation schemes are tested, as mentioned in Section \ref{IMGmethod}. For both models the Inverse Average Weighted interpolation technique is used as it yielded the best results in the classifiers performance. The average accuracies of different interpolation methods tested are presented in Table~\ref{tab:Interp} for SAD dataset with energy and entropy features.
 
Regarding the choice of the image size, different values of K pixels are considered in the analysis and the average accuracies of model 2 classification performance are represented in Table~\ref{tab:imgSize} for the CNN classifier for both datasets in subject-independent cases.

\begin{table}[!t]\centering
 \caption{Average Performance of Different Interpolation Methods}
\label{tab:Interp}
\begin{tabular}{cc}
 \toprule
Interpolation method     &Average recognition perf.(\%)  \\
 \midrule
 Nearest Neighbor  &   $75$ \\ 
 Bilinear &$76.56$    \\
  IDW with 0 BP  &$84.37$ \\
 Cubic B-spline   &$89.06$    \\
 IDW with NN BP  &$92.19$ \\
\bottomrule
\end{tabular}
\end{table}


\begin{table}[!t]\centering
 \caption{Average Performance using Different Image Sizes K$\times$K}
\label{tab:imgSize}
\begin{tabular}{llll}
 \toprule
Image size     &SAD dataset  &DEAP (Valence) &DEAP (Arousal)\\
 \midrule
 K=10  &  $90.65\%$ &   $88.28\%$ & $86.79\%$\\ 
 K=15 &$92.19\%$  &   $91.85\%$ &  $91.06\%$\\ K=20  &$89.06\%$ &   $92.17\%$ & $89.75\%$\\
 K=25  &$85.93\%$   &   $80.93\%$ & $82.16\%$\\
\bottomrule
\end{tabular}
\end{table}

It should be noted that for both datasets, a smaller number of channels was also used in the analysis. However using all 34 channels for SAD and 32 for DEAP gave superior results in terms of classification accuracies. 

\subsection{Comparison with EEGNet}
To further emphasize the significance of the results achieved by our model, the classification performances of both datasets were investigated using EEGNet. EEGNet is  compact convolutional neural network with successful generalized architecture for EEG-based brain–computer interfaces \cite{lawhern2018eegnet}. The first block in the network is a temporal convolution, followed by a depthwise convolution to learn frequency-specific spatial filters. The second block is a separable convolution block, which is a combination of a depthwise convolution and a pointwise convolution. The last block consists of dense layers and softmax classification at the end.
For performance analysis using EEGNet, cross-subject classification approaches are only considered. There are F1 2D convolutional filters of size (1,64), with filter length 64 chosen to be half the sampling rate of the data, which is 128Hz in all cases. For the depthwise convolution there are D filters of size (C, 1) to learn the spatial filters, where C is the number of channels. In the separable convolution block, F2 pointwise filters are used. Finally dropout rate is chosen to be 0.25, as the training set sizes are larger in cross-subject tasks compared to within-subject tasks. Different combinations of $F1=\{8,4,16\}$, $D=\{1,2,4\}$, and $F2=\{8,16,32,64,128\}$ are considered. The best average classification performances obtained for SAD, Valence and arousal in cross-subject analysis using EEGNet are $75.5\%$, $72.35\%$, and $74.76\%$ respectively. Since EEGNet is considered a successful generic EEG-based architecture, it was important to provide a comparison between the methods. 

\subsection{Effect of data augmentation}
The performance results listed above belong to classifiers with larger training sets, which were expanded using data augmentation. The significance of creating more robust model by increasing the training dataset is reflected in the higher classification accuracies of training with augmented samples achieved in both models. Wilcoxon Signed Rank test was conducted using augmentation as a factor, $p\textless{0.05}$ was achieved indicating the significance of data augmentation. The results are summarized in Table~\ref{tab:augM1} for model 1, and Table~\ref{tab:augM2} for model 2. In these tables, the second column represents performance of CNN using original training set, and the third column using the augmented training set.

\begin{table}[!t]
 \caption{Data Augmentation Affect Model 1 }
\label{tab:augM1}
\begin{tabular}{lll}
 \toprule
& Classification Accuracy(\%)& \\
  \midrule
  &no augmentation & with augmentation\\
 \midrule
 SAD  &$79.68$ & $84.38$\\
 Arousal   &$79.53$    &$83.94$\\
 Valence  &   $82.74$  & $84.77$ \\
\bottomrule
\end{tabular}
\end{table}


\begin{table}[!t]
 \caption{Data Augmentation Affect Model 2 }
\label{tab:augM2}
\begin{tabular}{lll}
 \toprule
& Classification Accuracy(\%)& \\
  \midrule
  &no augmentation & with augmentation\\
 \midrule
 SAD  &$89.06$ & $92.19$\\
 Arousal   &$87.27$    &$91.06$\\
 Valence  &   $87.15$  & $91.85$ \\
\bottomrule
\end{tabular}
\end{table}

\section{Conclusion}
In this paper, a new EEG-based classification approach was proposed. Unlike many other studies, this approach factored in the spatial configuration of the EEG sensors in signal analysis. A model that takes advantage of the knowledge of the locations of the electrodes was created to construct the EEG dataset. In order to assess the effectiveness of the model, two EEG datsets were used for analysis: SAD for patient/control classification and DEAP for emotion recognition. Overall results showed that the performance of various classifiers based on this model was $5\%$-$8\%$ higher in accuracy, compared with the same classifiers which ignored the configuration. In addition, use of the entropy along with the energy as relevant features in the EEG based classification task, reduced the error rate for the different classifiers (CNN, SVM, kNN) in both datasets.
It is also found that data augmentation for the training set is very important to further enhance the performance. This improvement is especially noticeable for model 2. The effectiveness of the proposed approach is reflected in the results which prove its superiority over other approaches, including those who also considered the spatial topology of the sensors in their analysis. The exceeding performance of model 2 is apparent, regardless of the classifier type, features, and most importantly, the nature of EEG dataset.

\bibliography{myRef.bib}
\bibliographystyle{IEEEtran}

\end{document}